\documentclass[11pt]{amsart}
\usepackage{latexsym,amssymb,amsmath,amscd,amsthm}
\numberwithin{equation}{section}
\theoremstyle{remark}

\newtheorem{example}{{\bf EXAMPLE}}[section]
\newcommand{\bq}{\begin{equation}}
\newcommand{\bea}{\begin{array}}
\newcommand{\eea}{\end{array}}

\newcommand{\ga}{\alpha}
\newcommand{\gep}{\epsilon}
\newcommand{\gD}{\Delta}
\newcommand{\gl}{\lambda}
\newcommand{\gL}{\Lambda}
\newcommand{\gb}{\beta}

\newcommand{\mf}{\mathfrak}

\newcommand{\ci}{\circ}
\newcommand{\ul}[1]{\underline{#1}}

\newcommand{\ola}{\overleftarrow}

\newcommand{\gG}{\Gamma}

\newcommand{\gs}{\sigma}

\newcommand{\gd}{\delta}
\newcommand{\pp}{\partial}

\newcommand{\ora}{\overrightarrow}

\newcommand{\tl}{\tilde}

\newcommand{\gk}{\kappa}

\newcommand {\nk}{\left(\begin{array}{c}
n\\
k\end{array}\right)}
\newcommand {\ms}{\left(\begin{array}{c}
m\\
s\end{array}\right)}

\newcommand{\bs}{\blacksquare}

\newcommand{{\DDD}}{D\!\!\!\!\!\!-}
\title{INTEGRABLE SYSTEMS AS QUANTUM MECHANICS}
\author{Robert Carroll\\University of Illinois, Urbana, IL 61801}
\date{September, 2003\thanks{email: rcarroll@math.uiuc.edu}}
\begin{document}

\bibliographystyle{plain}

\begin{abstract}
This is a mainly expository sketch showing how some integrable systems 
(e.g. KP or KdV) can be viewed
as quantum mechanical in nature.
\end{abstract}

\maketitle

%\newpage
%\tableofcontents
%\addcontentsline{toc}{section}{subsection}
%\setcounter{tocdepth}{3}

%\tableofcontents

%\abstract{We show how uncertainty can arise in a trajectory representation, refine
%some aspects of the enhanced dKdV theory related to $(X,\psi)$ duality, and discuss
%some features of the time dependent Hamilton-Jacobi theory.}

%\section{INTRODUCTION}
%\renewcommand{\theequation}{1.\arabic{equation}}
%\setcounter{equation}{0}

\section{INTRODUCTION}
\renewcommand{\theequation}{1.\arabic{equation}}
\setcounter{equation}{0}

We review briefly some ideas and constructions for dKP (from KP) and its
Hamiltonian dynamics, discuss its quantization via Moyal back to KP, and treat
the KP hierarchy itself as a sequence of quantum Hamiltonian systems based on the 
dKP dynamics.  Some connections to q-theories are also indicated.  The
constructions are all well known (cf.
\cite{b2,b1,ch,c1,c2,c3,c6,c7,c8,c9,c10,cq,c16, d1,f4,g1,h2,k2,k1,s1,s2,t2,z2})
but we try to organize matters and perhaps see them from a different perspective. 
We will use dKdV and KdV as a source of  simple examples.

\section{BACKGROUND}
\renewcommand{\theequation}{2.\arabic{equation}}
\setcounter{equation}{0}

There is an enormous literature on KP and KdV which we do not try to reference
(see e.g. \cite{ch,c14,d1}).  For a brief sketch we follow \cite{c8} and simply
write down the relevant formulas.  Thus one begins with two pseudodifferential
operators
${\bf (A1)}\,\,L=\pp+\sum_1^{\infty}u_{n+1}\pp^{-n}$ and
$W=1+\sum_1^{\infty}w_n\pp^{-n}$ (Lax and gauge operator respectively) with $L=W\pp
W^{-1}$.  Note here the generalized Leibniz rule
\bq\label{1}
\pp^nf=\sum_0^{\infty}{\nk}(\pp^kf)\pp^{n-k}
\end{equation}
The KP hierarchy is then determined by Lax equations ${\bf (A2)}\,\,\pp_nL=[B^n,L]=
B_nL-LB_n$ where $\pp_n\sim\pp/\pp t_n$ and $B_n=L_{+}^n$ is the differential part of
$L^n=L^n_{+}+L^n_{-}$.  One can also express this via the Sato equation ${\bf (A3)}\,\,
\pp_nW\,W^{-1}=-L^n_{-}$.  Now define wave functions via ${\bf (A4)}\,\,\psi=Wexp(\xi)=
w(t,\gl)exp(\xi)$ where $\xi=\sum_1^{\infty}t_n\gl^n$ and $w(t,\gl)=1+\sum_1^{\infty}
w_n(t)\gl^{-n}$ (with $t_1=x$).  There is also an adjoint wavefunction ${\bf (A5)}\,\,
\psi^*=W^{*-1}exp(-\xi)=w^*(t,\gl)exp(-\xi)$ where
$w^*(t,\gl)=1+\sum_1^{\infty}w_i^*(t)\gl^{-i}$; there are then equations
\bq\label{2}
L\psi=\gl\psi;\,\,\pp_n\psi=B_n\psi;\,\,L^*\psi=\gl\psi^*;\,\,\pp_n\psi^*=-B^*_n\psi^*
\end{equation}
Next one has the fundamental object, namely the tau function, which yields the wave
functions via vertex operators $X,\,X^*$ in the form
\bq\label{3}
\psi(t,\gl)=\frac{X(\gl)\tau(t)}{\tau(t)}=\frac{e^{\xi}\tau(t-[\gl^{-1}])}{\tau(t)};\,\,
\psi^*(t,\gl)=\frac{X^*(\gl)\tau(t)}{\tau(t)}=\frac{e^{-\xi}\tau(t+[\gl^{-1}])}
{\tau(t)}
\end{equation}
We write $\tau_{\pm}(t)=\tau(t\pm [\gl^{-1}])=exp(\pm\xi(\tl{\pp},\gl^{-1})
\tau$ where $\tl{\pp}=(\pp_1,(1/2)\pp_2,(1/3)\pp_3,\cdots)$ and $t\pm [\gl^{-1}]=
(t_1\pm\gl^{-1},t_2\pm(1/2)\gl^{-2},\cdots)$ and we can also specify
\bq\label{4}
e^{\xi}=exp\left(\sum_1^{\infty}t_n\gl^n\right)=\sum_0^{\infty}p_j(t_1,\cdots,t_j)\gl^j;
\,\,p_j(t)=\sum\left(\frac{t_1^{k_1}}{k_1!}\right)\left(\frac{t_2^{k_2}}{k_2!}\right)
\cdots
\end{equation}
where the $p_j$ are elementary Schur polynomials (note $\sum jk_j=n$).  One recalls
also the famous Hirota bilinear identity 
${\bf (A6)}\,\,\oint_{\infty}\psi(t,\gl)\psi^*(t',\gl)d\gl=0$ (residue integral
around $\infty$).  Using \eqref{3} this can be written as ${\bf (A7)}\,\,
\oint_{\infty}\tau(t-[\gl^{-1}])\tau(t'+[\gl^{-1}])exp[\xi(t,\gl)-\xi(t',\gl)]d\gl=0$
which leads to the characterization of the tau function via ($t\to t-y$ and $t'\to
t'+y$)
\bq\label{5}
\left(\sum_0^{\infty}p_n(-2y)p_{n+1}(\tl{D})exp\left(\sum_1^{\infty}y_iD_i\right)\right)
\tau\cdot\tau=0
\end{equation}
where $D_i$ is the Hirota derivative defined as ${\bf (A8)}\,\,D_j^ma\cdot
b=(\pp^m/\pp s_j^m)a(t_j+s_j)b(t_j-s_j)|_{s=0}$ and $\tl{D}=(D_1,(1/2)D_2,
(1/3)D_3,\cdots)$.  In particular from the coefficient of the free parameter $y_n$ in 
\eqref{5} one obtains ${\bf
(A9)}\,\,D_1D_n\tau\cdot\tau=2p_{n+1}(\tl{D})\tau\cdot\tau$ and these are called
Hirota bilinear equations.
Motivated via finite zone situations and Riemann surfaces (cf. \cite{c14}) where the
tau function is intimately related to theta functions one can express the Fay
trisecant identity in a form referred to as the Fay identity ${\bf (A10)}\,\,
\sum_{c.p.}(s_0-s_1)(s_2-s_3)\tau(t+[s_0]+[s_1])\tau(t+[s_2]+[s_3])=0$ where
c.p. means cyclic permutations (cf. \cite{a1,c14}).  Differentiating this in $s_0$,
setting $s_0=s_3=0$, dividing by $s_1s_2$, and shifting $t\to t-[s_2]$, leads to the
differential Fay identity ($\pp=\pp_x$)
\bq\label{6}
\tau(t)\pp\tau(t+[s_1]-[s_2])-\tau(t+[s_1]-[s_2])\pp\tau(t)=
\end{equation}
$$=(s_1^{-1}-s_2^{-1})
\{\tau(t+[s_1]-[s_2])\tau(t)-\tau(t+[s_1])\tau(t-[s_2])\}$$
The Hirota equations {\bf (A9)} can also be derived from \eqref{6} by taking the limit
$s_1\to s_2$.
\\[3mm]\indent
For the dispersionless theory (dKP) one thinks of fast and slow variables for example
taking $t_n\to\gep t_n=T_n$ and $t_1=x\to \gep x=X$.  Then letting $\gep\to 0$ the KP
equation ${\bf (A11)}\,\,u_t=(1/4)u_{xxx}+3uu_x+(3/4)\pp^{-1}u_{yy}$ ($u\sim u_2$) goes
to $\pp_TU=2UU_X+(3/4)\pp^{-1}U_{YY}$ where $\pp\sim\pp_X$ now (for a discussion of
the passage $u_n\to U_n$ we refer to \cite{ch,c10} - generally we think here of
$u_n(T/\gep)\sim U_n(T)+O(\gep)$ etc.).  Now take a WKB form for the wavefunction of
the form ${\bf (A12)}\,\,\psi=exp(S(T,\gl)/\gep)$ and define $P=\pp_XS$; then as
$\gep\to 0$ the equation $L\psi=\gl\psi$ becomes ${\bf (A13)}\,\,\gl=P+\sum_1^{\infty}
U_{n+1}P^{-n}$ with inverse $P=\gl-\sum_1^{\infty}P_{i+1}\gl^{-i}$.  Further from
$B_n\psi=\sum_0^nb_{nm}(\gep\pp)^m\psi$ one obtains ${\bf (A14)}\,\,\pp_nS={\mf
B}_n(P)=\gl_{+}^n$ where $B_n=L_{+}^n\to{\mf B}_n=\gl_{+}^n=\sum_0^nb_{nm}P^m$.
Consequently the KP hierarchy becomes ${\bf (A15)}\,\,\pp_nP=\pp{\mf B}_n$
(note $\pp_nS={\mf B}_n\Rightarrow\pp_nP=\pp{\mf B}_n$ and $\pp_n\sim\pp/\pp T_n$
here).  One can now write
\bq\label{7}
S=\sum_1^{\infty}T_n\gl^n-\sum_1^{\infty}\frac{\pp_mF}{m}\gl^{-m};\,\,
\tau=exp\left(\frac{1}{\gep^2}F(T)\right)
\end{equation}
and there results $P=\pp S={\mf B}_1$ with
\bq\label{8}
{\mf B}_n=\pp_nS=\gl^n-\sum_1^{\infty}\frac{F_{mn}}{m}\gl^{-m}
\end{equation}
Next following \cite{c9,t2} one can manipulate the differential Fay identity 
in various interesting ways and
in \cite{c8} the author and Y. Kodama derived a dispersionless limit of the Hirota
equations which provided polynomial identities among the coefficients
$F_{mn}=\pp_n\pp_mF$.  These dispersionless Hirota equations can be written in the
form ${\bf (A16)}\,\,F_{ij}=p_{j+1}(Z_1=0,Z_2,\cdots,Z_{j+1})$ where $p_n\sim$ Schur
polynomial and $Z_j=\sum_{m+n=j}(F_{mn}/mn)$.
One shows also that the $F_{mn}$ can be expressed as polynomials in the $P_{j+1}=
F_{1j}/j$.  There are many other results in \cite{c8} (including connections to 
D-bar techniques) and dKP has been developed and used in a number of fascinating
contexts in recent years (we refer e.g. to
\cite{b5,b3,b4,k3,k4,k5,k7,k6,m1,m2,m3,w1,z1}).
\\[3mm]\indent
{\bf REMARK 2.1.}
Another important aspect of dKP is the Hamiltonian theory originating in \cite{k2}
(cf. also \cite{ch,c6,c7,c8}).  It is convenient here to rescale the variables via
$t_n\to T'_n=nT_n$ with ${\mf Q}_n=(1/n){\mf B}_n$.  There results
\bq\label{9}
P'_n=\frac{dP}{dT'_n}=\pp{\mf Q}_n;\,\,X'_n=\frac{dX}{dT'_n}=-\pp_P{\mf Q}_n
\end{equation}
One can also show that (cf. \cite{c8})
\bq\label{10}
\frac{1}{P(\mu)-P(\gl)}=\sum_1^{\infty}\pp_P{\mf Q}_n(\gl)\mu^{-n}
\end{equation}
which is in fact equivalent to the dispersionless differential Fay identity
\bq\label{11}
\sum_{m,n=1}^{\infty}\mu^{-m}\gl^{-n}\frac{F_{mn}}{mn}=log\left(1-\sum
\frac{\mu^{-n}-\gl^{-n}}{\mu-\gl}\frac{F_{1n}}{n}\right)
\end{equation}
The kernel in \eqref{10} represents a Cauchy type kernel and following Kodama has a
version on Riemann surfaces related to the prime form.$\hfill\bs$

\section{INTEGRABLE SYSTEMS AND MOYAL}
\renewcommand{\theequation}{3.\arabic{equation}}
\setcounter{equation}{0}

For background we follow \cite{ch,c1,c2} and one recalls for wave functions
$\psi$ there are Wigner functions {\bf (WF)} given via
\bq
f(x,p)=\frac{1}{2\pi}\int dy\psi^*\left(x-\frac{\hbar}{2}y\right)exp(-iyp)
\psi\left(x+\frac{\hbar}{2}y\right)
\label{3.1}
\end{equation}
Then defining $f*g$ via 
\bq
f*g=f\, exp\left[\frac{i\hbar}{2}(\overleftarrow{\partial}_x
\overrightarrow{\partial}_p-\overleftarrow{\partial}_p\overrightarrow{\partial}_x
\right]\,g;
\label{3.2}
\end{equation}
$$f(x,p) * g(x,p)=f\left(x+\frac{i\hbar}{2}\overrightarrow{\partial}_p,
p-\frac{i\hbar}{2}\overrightarrow{\partial}_x\right)g(x,p)$$
time dependence of WF's is given by ($H\sim$ Hamiltonian)
\bq
\partial_tf(x,p,t)=\frac{1}{i\hbar}(H * f(x,p,t)-f(x,p,t) * H)=
\{H,f\}_M
\label{3.3}
\end{equation}
where $\{f,g\}_M\sim$ Moyal bracket.  As $\hbar\to 0$ this reduces to $\pp_tf-
\{H,f\}=0$ (standard Poisson bracket).  One can generalize and write out 
\eqref{3.2} in various ways.  For example replacing $i\hbar/2$ by $\gk$ one obtains as
in \cite{g1}
\bq
f*g=\sum_0^{\infty}\frac{\kappa^s}{s!}\sum_{j=0}^s(-1)^j{s \choose j}
(\partial_x^j\partial_p^{s-j}f)(\partial_x^{s-j}\partial_p^jg)
\label{3.4}
\end{equation}
leading to ($\{f,g\}_{\gk}=(f*g-g*f)/2\gk$)
\bq
\{f,g\}_{\kappa}=\sum_0^{\infty}\frac{\kappa^{2s}}{(2s+1)!}\sum_{j=0}^{2s+1}
(-1)^j{2s+1 \choose j}(\partial_x^j\partial_p^{2s+1-j}f)
(\partial_x^{2s+1-j}\partial_p^jg)
\label{3.5}
\end{equation}
which can also be utilized in the form
\bq\label{3.6}
f*g=fe^{\gk(\ola{\pp}_x\ora{\pp}_p-\ola{\pp}_p\ora{\pp}_x)}g=e^{[\gk(\pp_{x_1}\pp_{p_2}
-\pp_{x_2}\pp_{p_1})]}f(x_1,p_1)g(x_2,p_2)|_{(x,p)}=
\end{equation}
$$=\sum_0^{\infty}\frac{(-1)^r\gk^{r+s}}{r!s!}\frac{\pp^{r+s}f}{\pp x^r\pp p^s}\frac
{\pp^{r+s}g}{\pp p^r\pp x^s}=\sum_0^{\infty}\frac{\gk^n(-1)^{n-s}}{s!(n-s)!}
\left(\pp_x^{n-s}\pp_p^sf\right)\left(\pp_x^s\pp_p^{n-s}g\right)=$$
$$=\sum_0^{\infty}\frac{\gk^n}{n!}\sum_0^n(-1)^r\left(\pp_x^r\pp_p^{n-r}f\right)
\left(\pp_x^{n-r}\pp_p^rg\right)$$
Note e.g.
\bq\label{3.7}
g*f=g(x+\gk\pp_p,p-\gk\pp_x)f=f(x-\gk\pp_p,p+\gk\pp_x)g
\end{equation}
The Moyal bracket can then be defined via
\bq\label{3.8}
\{f,g\}_M=\frac{1}{\gk}\{f\,Sin[\gk(\ola{\pp}_x\ora{\pp}_p-\ola{\pp}_p\ora{\pp}_x)]g\}
=\frac{1}{2\gk}(f*g-g*f)=
\end{equation}
$$=\sum_0^{\infty}\frac{(-1)^s\gk^{2s}}{(2s+1)!}\sum_0^{2s+1}(-1)^j\left(
\begin{array}{c}
2s+1\\
j\end{array}\right)[\pp_x^j\pp_p^{2s+1-j}f][\pp_x^{2s+1-j}\pp_p^jg]$$
corresponding to $\gk\to i\gk$ in \eqref{3.12} below.
\\[3mm]\indent
Thus in \cite{g1} (cf. also \cite{s2}) one writes the Sato KP
hierarchy via ($v_{-2}=1,\,v_{-1}=0$)
\bq
\partial_mL=[L^m_{+},L]\,\,(m\geq 1);\,\,L=\sum_{-2}^{\infty}v_n
(\tilde{x})\partial_x^{-n-1}
\label{3.9}
\end{equation}
for $\tilde{x}=(x,t_2,\cdots$) while the Moyal KP hierarchy is written via
($u_{-2}=1,\,u_{-1}=0$)
\bq
\lambda\sim\Lambda=\sum_{-2}^{\infty}u_n(\tilde{x})P^{-n-1};
\,\,\partial_m\Lambda=\{\Lambda^m_{+},\Lambda\}_M\,\,(m\geq 1)
\label{3.10}
\end{equation}
where $\Lambda^m_{+}\sim (\Lambda^{*m})_{+}$ with
\bq
f*g=\sum_0^{\infty}\frac{\kappa^s}{s!}\sum_{j=0}^s(-1)^j{s \choose j}
(\partial_x^j\partial_P^{s-j}f)(\partial_x^{s-j}\partial_P^jg)
\label{3.11}
\end{equation}
leading to
\bq
\{f,g\}_{\kappa}=\sum_0^{\infty}\frac{\kappa^{2s}}{(2s+1)!}\sum_{j=0}^{2s+1}
(-1)^j{2s+1 \choose j}(\partial_x^j\partial_P^{2s+1-j}f)
(\partial_x^{2s+1-j}\partial_P^jg)
\label{3.12}
\end{equation}
Note $lim_{\kappa\to 0}\{f,g\}_{\kappa}=\{f,g\}=f_{\lambda}g_x-
f_xg_{\lambda}$ so $(KP)_M\to dKP$ as $\kappa\to 0$, namely
$\partial_m\Lambda=\{\Lambda^m_{+},\Lambda\}$ with $\Lambda^m\sim
\Lambda\cdots\Lambda$.  The isomorphism between $(KP)_{Sato}$ and $(KP)_M$ 
is then determined by relating $v_n$ and $u_n$ in the form $(\kappa=1/2$)
\bq
u_n=\sum_0^n2^{-j}{n \choose j}v^j_{n-j}
\label{3.13}
\end{equation}
where $n=0,1,\cdots$ and $v^j=\partial_x^jv_0$ (see Remark 3.1 for enhancement).
\\[3mm]\indent 
{\bf REMARK 3.1.}
This can be further clarified as follows.  First
we recall an important paper
\cite{f4} where one considers star products of the form
${\bf (B1)}\,\,f\star g=fg+\sum_{n\geq 1}h^nB_n(f,g)$ with
bilinear differential operators $B_n$.  In particular in \cite{f4} one shows that
any bracket of the form
\bq\label{3.14}
\{f,g\}=\sum_{r=1}^{\infty}\sum_{s=1}^{\infty}\gl^{r+s-2}\sum_{j=0}^r\sum_{k=0}^s
b_{rj,sk}(\pp_x^j\pp_y^{r-j}f)(\pp_x^k\pp_y^{s-k}g)
\end{equation}
may be transformed to one with $b_{00,10}=b_{00,11}=0$ and any such bracket satisfying
the Jacobi identity must be of the form
\bq\label{3.15}
\{f,g\}=\sum_{r=1}^{\infty}\gl^{r-1}\sum_{j=0}^r\sum_{k=0}^sb_{rjk}(\pp_x^j\pp_y^{r-j}
f)(\pp_x^k\pp_y^{r-k}g)
\end{equation}
By suitable calculation one shows also that \eqref{3.15} plus Jacobi is equivalent to
Moyal.  Note that the Jacobi condition for $\{f,g\}=(1/h)(f\star g-g\star f)$ can be
proved directly via associativity of $\star$ (exercise).  
Thus ${\bf (B2)}\,\,
\{\{f,g\},h\}+\{\{h,f\},g\}+\{\{g,h\},f\}=0$.
Now to connect the dKP theory with bracket \eqref{3.12} to a $\gk-KP$ theory with PSDO
bracket consider the PSDO symbol bracket 
${\bf (B3)}\,\,A\ci B=\sum(1/k!)\pp_{\xi}^kA(x,\xi)\pp_x^kB(x,\xi)$ 
(cf. \cite{ch,t5})
where $A\sim \sum a_i(x)\xi^i,\,\,\pp_{\xi}^kA=\sum a_i(x)\pp_{\xi}^k\xi^i$, and 
$\pp_x^kA=\sum \pp_x^ka_i(x)\xi^i$.  Note also
\bq\label{3.16}
A\ci_{\gk}B=Ae^{\gk\ola{\pp}_{\xi}\ora{\pp}_x}B=\sum\frac{\gk^n}{n!}\pp_{\xi}^nA
\pp_x^nB
\end{equation} 
and the bracket based on this; thus 
\bq\label{3.17}
A\ci_{\gk}B=A(x, \xi+\gk\pp_x)B(x,\xi);\,\,B\ci_{\gk}A=B(x,\xi+\gk\pp_x)A(x,\xi)
\end{equation}
and ${\bf (B4)}\,\,(1/\gk)A\ci_{\gk}B-B\ci_{\gk}A)=\{A,B\}_{\gk}$ is of
the form \eqref{3.15} with $b_{rr0}\ne 0,\,\,b_{r0r}\ne 0$, and all other coefficients
equal 0.  Also $b_{110}=-b_{101}$ and the Jacobi identity will follow from
associativity so in fact any bracket such as {\bf (B4)} is equivalent to Moyal in the
symbols involved ($\gk$ is arbitrary).  
Note here that associativity is not obvious but is proved in
\cite{k9} (note the $\gk$ can be absorbed in $\xi$ by rescaling).  The trick is to use
the formula
\bq\label{3.18}
(a\xi^n)(b\xi^r)=a\sum_{k\geq 0}\binom{n}{k}\pp_x^kb\xi^{n-k+r}
\end{equation}
which shows that (for $\gk=1$)
\bq\label{3.19}
A\ci_{\gk}B=\sum\frac{1}{m!}\pp_{\xi}^mA\pp_x^nB=\sum\frac{1}{m!}\cdot
\end{equation}
$$\cdot\sum
a_nn(n-1)\cdots(n-m+1)\xi^{n-m}\cdot\sum b_j^{(m)}\xi^j=\sum a_n\binom{n}{m}b_j^
{(m)}\xi^{n-m+j}$$
Now for associativity one checks that ${\bf
(B5)}\,\,[\xi^n(a\xi^r)]b=\xi^n[(a\xi^n)b]$ and we refer to \cite{k9} for further
details.$\hfill\bs$
\\[3mm]\indent
We recall also the standard symbol calculus for PSDO following e.g.
\cite{g1,m4,t5} (cf. {\bf (B3)}, \eqref{3.16}, \eqref{3.17}).
First one recalls from \cite{g1} the ring or algebra 
$\mf{A}$ of pseudodifferential
operators (PSDO) via PSD symbols (cf. also \cite{t5} for a more mathematical
discussion).  Thus one looks at formal series ${\bf
(B6)}\,\,A(x,\xi)=\sum_{-\infty}^na_i(x)\xi^i$ where $\xi$ is the symbol for
$\pp_x$ and $a_i(x)\in{\bf C}^{\infty}$ (say on the line or circle).  The
multiplication law is given via the Leibnitz rule for symbols ${\bf
(B7)}\,\,A(x,\xi)\ci B(x,\xi)=\sum_{k\geq 0}(1/k!)A^k_{\xi}(x,\xi)B_x^{(k)}(x,\xi)$
where $A_{\xi}^k(x,\xi)=\sum_{-\infty}^na_i(x)(\xi^i)^{(k)}$ and $B_x^{(k)}(x,\xi)=\sum_
{-\infty}^nb_i^{(k)}(x)\xi^i$ with $b_i^{(k)}(x)=\pp_x^kb_i(x)$.  This gives a
Lie algebra structure on $\mf{A}$ via ${\bf (B8)}\,\,[A,B]=A\ci B-B\ci A$.  Now let
A be a first order formal PSDO of the form ${\bf
(B9)}\,\,A=\pp_x+\sum_{-\infty}^{-1}a_i(\tl{x})\pp_x^i$ where $\tl{x}\sim (x,t_2,
t_3,\cdots)$.  Then the KP hierarchy can be written in the form ${\bf (B10)}\,\,(\pp
A/\pp t_m)=[(A^m)_{+},A]$ which is equivalent to a system of evolution equations
${\bf (B11)}\,\,(\pp a_i/\pp t_m)=f_i$ where the $f_i$ are certain universal
differential polyomials in the $a_i$, homogeneous of weight $m+|i|+1$ where
$a^j_{-i}$ has weight $|i|+j+1$ for $a^j\sim\pp_x^ja$.
Somewhat more traditionally (following \cite{t5} - modulo notation and various
necessary analytical details), one can write
\bq\label{3.20}
Au(x)=(2\pi)^{-1}\int e^{ix\cdot\xi}a(x,\xi)\hat{u}(\xi)d\xi
\end{equation}
where $\hat{u}(\xi)=\int exp(-ix\cdot \xi)u(x)dx$.  One takes $D=(1/i)\pp_x$ and
writes $a=symb(A)$ with $A=op(a)\sim\dot{A}$ where the $\cdot$ is to mod out
$\mf{S}^{-\infty}$ (we will not be fussy about this and will simply use A).  The
symbol of $A\ci B$ is then formally
\bq\label{3.21}
(a\odot b)(x,\xi)=\sum\frac{1}{\ga!}\pp_{\xi}^{\ga}a(x,\xi)D_x^{\ga}b(x,\xi)
\end{equation}
corresponding to {\bf (B7)}, while $[A,B]=AB-BA$ corresponds to the symbol
${\bf (B12)}\,\,\{a,b\}=(\pp a/\pp\xi)(\pp b/\pp x)-(\pp a/\pp x)(\pp b/\pp \xi)$
(note $\widehat{P(D)T}=P(\xi)\hat{T}$).
In any event it is clear that the algebra of differential operators on a manifold M
(which we have sometimes loosely referred to as quantum operators) may be considered as
a noncommutative deformation of the algebra of functions on $T^*M$ and the extension to
PSDO brings one into the arena of integrable systems.

\section{QUANTUM MECHANICS}
\renewcommand{\theequation}{4.\arabic{equation}}
\setcounter{equation}{0}

We refer now to \cite{z2} where a lovely survey appears
(cf. also \cite{ch,c4,c5,c17,f1}).  There are three logically autonomous alternative
paths to quantization, namely Hilbert space operators, path integrals, and deformation
quantization.  In fact the Wigner-Weyl-Moyal formulation gives a complete coverage
as follows.  It is based on the Wigner function (WF) which is a quasi probabiity
distribution in phase space defined via 
\bq\label{4.6}
f(x,p)=\frac{1}{2\pi}\int dy\psi^*(\left(x-\frac{\hbar}{2}y\right)e^{-iyp}
\psi\left(x+\frac{\hbar}{2}y\right)
\end{equation}
Here one has ${\bf (C1)}\,\,\int dpdxf(x,p)=1$ and in the classical limit $\hbar\to
0$
$f$ reduces to the probability density in coordinate space x (usually highly localized)
multiplied by delta functions in momentum; thus the classical limit is ``spiked".
The WF is manifestly real and constrained by the Schwartz inequality to be bounded with
$-(2/\hbar)\leq f(x,p)\leq (2/\hbar)$ (the bound disappearing in the spikey classical
limit).  Projection in $x$ or $p$ leads to marginal probability densities, namely, a
spacelike shadow $\int dpf(x,p\rho(x)$ or a momentum space shadow $\int
dxf(x,p)=\gs(p)$.   The WF can and most frequently does become negative in some
regions of phase space.  Nevertheless WF is a distribution function
providing the integration measure in phase space which yields expectation values from 
phase space c-number functions.  Such functions can be classical functions but in
general are associated to suitably ordered operators via Weyl's correspondence rule.
Given such an operator ordered via
\bq\label{4.8}
{\mf G}({\mf x},{\mf p})=\frac{1}{(2\pi)^2}\int d\tau d\gs dxdp\,g(x,p)e^{i\tau({\mf p}
-p)+i\gs({\mf x}-x)}
\end{equation}
the corresponding phase space function $g(x,p)$ (classical kernel) is obtained via
${\bf (C2)}\,\,{\mf p}\to p$ and ${\mf x}\to x$.  The operators expectation value is
then a phase space average ${\bf (C3)}\,\,<{\mf G}>=\int dxdp\,f(x,p)g(x,p)$.  The
classical kernel is often the unmodified classical expression such as $H=(p^2/2m)
+V(x)$ but it contains $\hbar$ when there are ordering ambiguities (see below).
This operation corresponds to tracing with a density matrix as indicated below.
\\[3mm]\indent
The dynamical evolution is specified by Moyal's equation which extends the Liouville
theorem of classical mechanics (CM), namely $\pp_tf+\{f,H\}=0$, and is given by
${\bf (C4)}\,\,\pp_tf=(1/i\hbar)[H\star f-f\star H]$ where ${\bf (C5)}\,\,
\star\sim exp[(i\hbar/2)(\ola{\pp}_x\ora{\pp}_p-\ola{\pp}_p\ora{\pp}_x)$.
The right side of {\bf (C4)} is of course the Moyal bracket and is the essentially
unique 1-parameter associative deformation of the Poisson bracket (cf. \cite{f4,p1}).
In practice evaluation can be expressed through Bopp operators in the form
\bq\label{4.9}
f(x,p)\star g(x,p)=f\left(x+\frac{i\hbar}{2}\ora{\pp}_p,p-\frac{i\hbar}{2}\ora{\pp}_x
\right)g(x,p)
\end{equation}
The equivalent Fourier representation of the star product can be expressed via
\bq\label{4.11}
f\star g=\frac{1}{(\hbar\pi)^2}\int dudvdwdz\,f(x+u,p+v)g(x+w,p+z)e^{(2i/\hbar)
(uz-vw)}
\end{equation}
which exhibits noncommutativity and associativity.  There is a complete isomorphism
between star multiplication and operator multiplication indicated in
\bq\label{4.12}
{\mf A}{\mf B}=\frac{1}{(2\pi)^2}\int d\tau d\gs dxdp(a\star b)e^{i\tau({\mf p}-p)
+i\gs ({\mf x}-x)}
\end{equation}
One sees also from \eqref{4.11} that ${\bf (C6)}\,\,\int dpdx\,f\star g=\int dpdx\,
fg=\int dpdx\,g\star f$.  Note that the Moyal equation is necessary but does not
suffice to specify the WF for a system (e.g. $f(H)$ commutes with H).
\\[3mm]\indent
Static or stationary WF's obey more powerful stargenvalue equations (cf. 
\cite{c4,c5,f1})
\bq\label{4.13}
H(x,p)\star f(x,p)=H\left(x+\frac{i\hbar}{2}\ora{\pp}_p,p-\frac{i\hbar}{2}\ora{\pp}_x
\right)f(x,p)=f(x,p)\star H(x,p)=Ef(x,p)
\end{equation}
where $H\psi=E\psi$ and this amounts to a complete characterization of the WF's.
Indeed, using a simple Hamiltonian $p^2/2m +V(x)$ (without essential loss of
generality) one proves (cf.
\cite{z2}) that for real
$f(x,p)$ the Wigner formula
\eqref{4.6} for pure stationary eigenstates is equivalent to compliance with the 
stargenvalue equations \eqref{4.13} along with $f\star H=Ef$.  Conversely
the pair of stargenvalue equations for $f(x,p)=\int dy exp(-ipy)\tl{f}(x,y)$ leads
to $\tl{f}=(1/2\pi)\psi^*(x-(\hbar/2)y)\psi(x+(\hbar/2)y)$.
There are also a number of special properties for pure state f.  Thus from ${\bf
(C7)}\,\, f\star H\star g=E_ff\star g=E_gf\star g$ so for $E_g\ne E_f$ one has ${\bf
(C8)}\,\,f\star g=0$. Moreover for $f=g$ one gets then ${\bf (C9)}\,\,f\star H\star
f=E_ff\star f=H\star f\star f$ so ${\bf (C10)}\,\, f\star f\propto f$ and in fact
${\bf (C11)}\,\,f_a\star f_b=(1/\hbar)\gd_{ab}f_a$.  Here the normalization is
important since it prevents superposition (which is handled differently as in the
density matrix formulation).  Note also by virtue of {\bf (C6)} for different 
stargenfunctions one has ${\bf (C12)}\,\,\int dpdx\,fg=0$ so one must go negative
to offset positive overlap (a virtue of negativity).  Further note that integrating
\eqref{4.13} yields the expectation of the energy ${\bf (C13)}\,\,\int H(x,p)f(x,p)
dxdp=E\int f\,dxdp=E$ and from {\bf (C11)} we get ${\bf (C14)}\,\,\int f^2dxdp=
1/\hbar$.
\\[3mm]\indent
Next note ${\bf (C15)}\,\,<g^*\star g>\geq 0$ which leads to the uncertainty
principle. Indeed 
\bq\label{4.14}
\int dpdx(g^*\star g)=\hbar\int dxdp(g^*\star g)(f\star f)=\hbar\int dxdp(f\star
g^*)\star (g\star f)=\hbar\int dxdp|g\star f|^2
\end{equation}
To produce Heisenberg's uncertainty principle one chooses ${\bf (C16)}\,\,
g=a+bx+cp$ for arbitrary constants $a,b,c\in {\bf C}$.  The resulting positive semi-
definite form is then ${\bf (C17)}\,\,a^*a+b^*b<x\star x>+c^*c<p\star p>+(a^*b+b^*a)
<x>+(a^*c+c^*a)<p>+c^*b<p\star x>+b^*c<x\star p>\geq 0$.  The eigenvalues of the
corresponding matrix
are then non-negative and so must be the determinent.  Some calculation then leads to
${\bf (C18)}\,\,\gD x\gD p\geq \hbar/2$ (cf. \cite{z2} and see \cite{cq} for details
in calculation).
\\[3mm]\indent
Going now to time evolution the Moyal equation {\bf (C4)} is formally solved by
virtue of associative combinatoric operations completely analogous to Hilbert space
quantum mechanics (QM) through definition of a star-unitary evolution operator
(star exponential) in the form
\bq\label{4.17}
U_*(x,p,t)=e_*^{itH/\hbar}=1+(it/\hbar)H(x,p)+\frac{(it/\hbar)^2}{2!}H\star H +
\frac{(it/\hbar)^3}{3!}H\star H\star H+\cdots
\end{equation}
Given the WF at $t=0$ the solution to the Moyal equation is then ${\bf (C19)}\,\,
f(x,p,t)=U_*^{-1}(x,p,t)\star f(x,p,0)\star U_*(x,p,t)$.  For variables $x,p$ this
collapses to classical trajectories
\bq\label{4.18}
\dot{x}=\frac{x\star H-H\star x}{i\hbar}=\pp_pH;\,\,\dot{p}=\frac{p\star H-
H\star p}{i\hbar}=-\pp_xH
\end{equation}
Thus for the harmonic oscillator ${\bf (C20)}\,\,x(t)=xCos(t)+pSin(t)$ and $p(t)=
pCos(t)-xSin(t)$ so the functional form of the WF is preserved along classical phase
space trajectories via ${\bf (C21)}\,\,f(x,p,t)=f(xCos(t)-pSin(t),pCos(t)+xSin
(t),0)$.

\section{MOMENTUM CALCULI}
\renewcommand{\theequation}{5.\arabic{equation}}
\setcounter{equation}{0}

We summarize first some features of the basic situation.
\begin{example}
Thus go back to Remark 2.1 and the Hamiltonian theory for dKP in the form \eqref{9},
written here as ${\bf (D1)}\,\,\dot{P}=\pp{\mf Q}_n$ and $\dot{X}_n=-\pp_P{\mf Q}_n$.
We know KP is equivalent to MdKp (Moyal dKP) and examine here the nature of
quantizing {\bf (D1)} via Moyal-Wigner-Weyl (MWW) formulas.  In {\bf (D1)} the T
variable is $\tau_n=T'_n=nT_n$ where $\gep t_n\sim T_n$ and we recall here the origin
of  {\bf (D1)} from \cite{c7,c8,k2}.  Thus as in Section 2 one arrives at
$\pp_nS={\mf B}_n$ and we set 
$n{\mf Q}_n={\mf B}_n$.  Then rescaling $\tau_n=T'_n=nT_n$ and writing $\pp_n\sim
\pp/\pp\tau_n$ now we have (for $P=\pp_XS$ and ${\mf B}_n={\mf B}_n(P,X)$)
${\bf (D2)}\,\,\pp_nP=\pp_X{\mf Q}_n=\pp{\mf Q}_n+\pp_P{\mf Q}_n\pp P$ (in an obvious
notation).  Then thinking of $P=P(X,\tau)$ write 
\bq\label{4.23}
\dot{P}_n=\pp_nP+\pp P\dot{X}_n=\pp{\mf Q}_n+\pp_P{\mf Q}_n\pp P+\pp P\dot{X}_n
\end{equation}
This is then incorporated into a HJ theory with S as a generating function and
equations {\bf (D1)}.  Now the quantization of the dynamical system via classical MWW
methods involves Hamiltonians $H(X,P)={\mf Q}_n(X,P)$ and equations (using $\tau$ as
the evolution time and $(X,P)\sim (x,p)$) ${\bf (D3)}\,\,i\hbar\dot{X}_n=\{X,{\mf
Q}_n\}_M$ and ${\bf (D4)}\,\,i\hbar\dot{P}_n=\{P,{\mf Q}_n\}_M$ where
$\{f,g\}_M=f\star g- g\star f$ with $\star$ as in {\bf (C5)}.  Hence as in
\eqref{4.18} we have ${\bf (D5)}\,\,\dot{X}_n=\pp_P{\mf Q}_n$ and
$\dot{P}_n=-\pp_X{\mf Q}_n$ repeated (i.e. the ``motion" is along the ``classical"
trajectories). One can ask now what this means in the standard KP theory.  Recall
from Remark 3.1 how the symbol calculus for PSDO is equivalent to Moyal dKP so in
some sense we will have a Hamiltonian theory in the symbols of KP operators (based
perhaps on some sort of relation such as \eqref{3.12}).$\hfill\bs$
\end{example}
\indent
Now we look at momentum calculi
(cf. also \cite{b2,d1,t1}).  The approach of \cite{b2} is
based on Lie algebras, Poisson structures, and R-matrices; the demands of the general
framework adopted however seem to restrict severely the range of applicability (e.g.
KP is not included).  Thus we mainly omit this.
Another, more flexible, approach is developed in \cite{d1} (cf. also \cite{t1,t9})
One begins with the standard phase
space star product
\bq\label{4.34}
A(x,p)*B(x,p)=e^{\gk(\pp_x\pp_{\tl{p}}-\pp_p\pp_{\tl{x}})}A(x,p)B(\tl{x},\tl{p})|_
{\tl{x},\tl{p}=(x,p)}
\end{equation}
with conventional Moyal bracket ${\bf
(D6)}\,\,\{A(x,p),B(x,p)\}_{\gk}=(1/2\gk)(A*B-B*A)$.  As usual one has $lim_{\gk\to
0}\{A,B\}_{\gk}=\{A,B\}$ (Poisson bracket).  The star product gives the momentum an
operator character via (note $\hbar/2\sim\gk$ when comparing notations)
\bq\label{4.35}
p^n*p^m=p^{m+n};\,\,p^n*f(x)=\sum_0^n\binom{n}{m}(2\gk)^mf^{(m)}*p^{n-m};
\end{equation}
$$\binom{n}{m}=\frac{n(n-1)\cdots (n-m+1)}{m!};\,\,\binom{n}{0}=1$$
Up to normalization these are precisely the relations satisfied by the derivative
operator.  
Let us check here some calculations based on \eqref{3.6} where
\bq\label{4.36}
f\star g=fe^{\gk(\ola{\pp}_x\ora{\pp}_p-\ola{\pp}_p\ora{\pp}_x)}g=
\sum_0^{\infty}\frac{\gk^s}{s!}\sum_0^s(-1)^j(\pp_x^j\pp_p^{s-j}f)(\pp_x^{s-j}\pp_p^jg)
\end{equation}
\bq\label{4.37}
p^m\star g=\sum\frac{\gk^s}{s!}\pp_p^sp^m\pp_x^sg=\sum_0^m{\ms}\gk^sp^{m-s}g^{(s)};
\end{equation}
$$f\star p^m=\sum\frac{\gk^s}{s!}(-1)^s\pp_x^sf\pp_p^sp^m=\sum_0^m{\ms}(-\gk)^sp^{m-s}
f^{(s)};$$
$$g^{(s)}\star p^{m-s}=\sum_0^{m-s}(-\gk)^jp^{m-s-j}f^{(s+j)}$$
Look at some low order terms $1\star f=f,\,\,p\star f=pf-\gk f',\,\,f\star p=pf-\gk
f',$ and $f\star 1 =f$.  Then $p\star f=pf=\gk f'=f\star p+2\gk f'=f\star
p+(2\gk)(f'\star 1)$ as in \eqref{4.35}.  For $p^2$ we have $p^2\star f=p^2f+
\left(\begin{array}{c}
2\\
1\end{array}\right)\gk pf'+\gk^2f''$ with $f\star p^2=p^2f-\left(\begin{array}{c}
2\\
1\end{array}\right)\gk pf'+\gk^2f''$.  Hence $p^2\star f=f\star p^2+2\gk
\left(\begin{array}{c}
2\\
1\end{array}\right)\gk pf'=f\star  p^2+2\gk\left(\begin{array}{c}
2\\
1\end{array}\right)
[f'\star p+\gk f'']=f\star p^2+(2\gk)\left(\begin{array}{c}
2\\
1\end{array}\right)(f'\star p)+(2\gk)^2f''\star 1=\sum_0^2\left(\begin{array}{c}
2\\
m\end{array}\right)
(2\gk)^mf^{(m)}\star p^{n-m}$.  Thus \eqref{4.35} seems reasonable and to compare with
derivation operators note also $(\gk\pp f)^2f=\gk^2(f''g+2f'g'+fg'')$ and
$(f'\gk\pp)g=f'\gk g'$ so $4(\gk\pp)^2f\sim f(2\gk\pp)^2+4\gk f'(2\gk\pp)+(2\gk)^2f''$.
Thus one has
\bq\label{4.38}
(2\gk\pp)^nf=\sum_0^n{\nk}(2\gk)^kf^{(k)}(2\gk\pp)^{n-k}
\end{equation}
upon extrapolation.
\\[3mm]\indent
Now one defines two classes of Lax operators on the phase space via
\bq\label{4.39}
L_n=p^n+u_1(x)*p^{n-1}+u_2(x)*p^{n-2}+\cdots+u_n(x);
\end{equation}
$$\gL_n=p^n+u_1(x)*p^{n-1}
+\cdots+ u_n(x)+u_{n+1}(x)*p^{-1}+\cdots$$
Thus one has replaced the space of pseudodifferential operators by that of
polynomials in momentum which inherits an operator structure through the star product
and defines an algebra.  This will be called the momentum algebra $M_n$ and one
notices that this is different from the concept of pseudodifferential operators 
(PSDO) with
the coefficients taken from the Moyal algebra of \cite{t10}.  All of the properties
of PSDO carry through with suitable redefinitions.  In particular thinking of the
residue as the coefficient of the $p^{-1}$ term one gets ${\bf (D7)}\,\,Res
\{A,B\}_{\gk}=(\pp_xC)$ exhibiting the residue as a total derivative.  Consequently
one can define ${\bf (D8)}\,\,Tr(A)=\int dx\,Res(A)$ which is unique (with the usual
assumptions of asymptotic decrease) and satisfies cyclicity.  For a general Lax
operator $\gL_n$ one checks immediately that 
\bq\label{4.40}
\frac{\pp\gL_n}{\pp t_k}=\left\{\gL_n,\left(\gL_n^{k/n}\right)_{\geq m}\right\}_{\gk};
\,\,\,(k\ne \ell n)
\end{equation}
defines a consistent Lax equation provided $m=0,1,2$ and the projectors are defined
with respect to the star product (the differencein ordering here can be adjusted via
$t_k\to -t_k$ if desired).
Note here $\gL^{k/n}=\gL^{1/n}*\cdots*\gL^{1/n}$
with k factors with the $n^{th}$ root determined recursively.  The projection with
$m=0$ is denoted by $(\,\,)_{+}$ and will be referred to as the standard Moyal-Lax
representation (the others are called nonstandard and are not considered here).  Note
that
${\bf (D9)}\,\,lim_{\gk\to 0}(\gL*\gL')_{+}=(\gL\gL')_{+}$ where the
factors on the right are phase space functions (not operators).  Thus one can go the Lax
representation of the dispersionless limit in a natural manner (cf. \cite{c7,c8,t1}). 
In fact in this limit one gets ${\bf (D10)}\,\,\pp_t\gL_n=\{\gL_n,(\gL_n^{k/n})_{+}$
with the standard Poisson bracket.  Further one determines conserved charges via
${\bf (D11)}\,\, H_k=Tr\,\gL_n^{k/n}\,\,\,(k\ne \ell n)$, proves that different flows
commute, and defines Hamiltonian structures in a straightforward manner.
For illustration consider
\begin{example}
In \cite{d1} the KdV hierarchy is developed via the Lax operator $L=p^2+u(x)$ where
${\bf (D12)}\,\,(L^{3/2})_{+}=p^3+(3/2)u*p+(3\gk/2)u^{(1)}$ ($u^{(1)}=\pp_xu$ - we
have changed the coefficient of $u^{(1)}$ to agree with calculations below in {\bf
(D25)}). 
One gets then
\bq\label{4.41}
\pp_tL=\left\{L,\left(L^{3/2}\right)_{+}\right\}_{\gk}\Rightarrow \pp_tu=
-\left(\gk u^3+\frac{3}{2}uu^{(1)}\right)
\end{equation}
The first few conserved quantities are (not checked)
\bq\label{4.42}
H_1=Tr(L^{1/2})=\int dx(u/2);\,\,H_2=Tr(L^{3/2})=\int dx(u^2/4);
\end{equation}
$$H_3=Tr(L^{5/2})=\int dx(4\gk u^{(2)}u+u^3)$$
The commutativity of flows follows directly from the Moyal-Lax representation. 
$\hfill\bs$
\end{example}
\begin{example}
The conventional Lax equation in standard representation ${\bf (D13)}\,\,
\pp_{t_k}=[(L^{k/n})_{+},L]$ (L a PSDO) resembles a Hamiltonian equation with
$(L^{k/n})_{+}$ as a Hamiltonian.  However such a relation cannot be further
developed in the language of PSDO.  In contrast consider the Moyal-Lax representation
with an arbitrary flow in the
KdV hierarchy described by
${\bf (D14)}\,\,\pp_tL=\{L,(L^{(2n+1)/2})_{+}\}_{\gk}$ (L as in \eqref{4.36}).  Then
consider an action of the form ${\bf (D15)}\,\,S=\int
dt(p*\dot{x}-(L^{(2n+1)/2})_{+})$.  It is important to remember here that $L=L(p,x)$
but does not depend on time explicitly.  Thus one can think of $(L^{(2n+1)/2})_{+}$ as
the Hamiltonian on the phase space.  That this is valid follows from the
Euler-Lagrange equations ($(2n+1)/2=\ga$ - cf. also Example 4.1)
\bq\label{4.43}
\dot{x}=\frac{\pp(L^{\ga})_{+}}{\pp p}=\{x,(L^{\ga})_{+}\}_{\gk};\,\,
\dot{p}=-\frac{\pp(L^{\ga})_{+}}{\pp x}=\{p,(L^{\ga})_{+}\}_{\gk}
\end{equation}
Further since L is a function on the phase space one has ${\bf
(D16)}\,\,\pp_tL=\{L,(L^{(2n+1)/2})_{+}\}_{\gk}$ so that the Moyal-Lax equation is
indeed a Hamiltonian equation with $L^{\ga}_{+}$ playing the role of Hamiltonian.
This procedure also goes through for the nonstandard representations.
$\hfill\bs$
\end{example}
\indent
The Moyal-Lax representation has now an advantage in that one can go to the
dispersionless limit of an integrable system by simply taking the limit $\gk\to 0$
and, while the Lax representations for various dispersionless integrable models are
known (cf. \cite{ch,c7,c8}), the determination of Hamiltonian structures
(at least the second) from such a Lax representation has often been open.  The
Moyal-Lax representation provides a solution to this problem in a natural way
(cf. \cite{c1,d1} for details). 
\\[3mm]\indent
{\bf REMARK 5.1.}
An approach similar to \cite{d1} was developed in \cite{t1} (cf. also \cite{t9})
and we sketch a few points here.
One considers an algebra of Laurent series of the form
$\gL=\{A;\,A=\sum_{-\infty}^Na_ip^i\}$ with coefficients depending on
$t_1=x,t_2,\cdots$.  $\gL$ can be decomposed as $\gL=\gL_{\geq k}\oplus\gL_{<k}$
for $k=0,1,\cdots$ where e.g. $\gL_{\geq k}=\{A=\sum_{i\geq k}a_ip^i\}$; the notation 
$\gL_{+}\sim \gL_{\geq 0}$ is used as before.  Evidently $\gL$ is an associative
but noncommutative algebra under the Moyal star product and one defines
$Res(A)=a_{-1}$ with trace as $Tr(A)=\int Res(A)$.  There results
\bq\label{4.44}
\int Res(A*B)=\int\sum_{i,j}\frac{\gk^{i-j+1}i!}{(i-j+1)!(j-1)!}(a_ib_j)^{(i-j+1)}=
\sum_i\int a_ib_{i+1}
\end{equation}
One notes that this is the same as in the dispersionless limit $\gk\to 0$ and because
of this the Hamiltonian formulation for the Moyal KdV (for example) becomes possible.
Indeed using \eqref{4.44} one shows that $Tr\{A,B\}_{\gk}=0$ and
$Tr(A*\{B,C\}_{\gk})=Tr(\{A,B\}_{\gk}*C)$.  To see this use \eqref{4.41} to replace
the Moyal star product within the trace by the ordinary multiplication.  Now given a
functional $F(A)=\int f(a)$ one defines a gradient via ${\bf
(D17)}\,\,d_AF=\sum_i(\gd f/\gd a_i)p^{-i-1}$ where the variational derivative is
defined via ${\bf (D18)}\,\,\gd f/\gd a_k=\sum_i(-1)^i(\pp^i\cdot(\pp f/\pp
a_k^{(i)})$ where $a_k^{(i)}=(\pp^i\cdot a_k)$ and $\pp\sim\pp/\pp x$.  Note
$\pp\cdot f=f'=\pp f/\pp x$ and $\pp f =f\pp+f'$.  The Moyal KdV hierarchy is defined
by the Lax equations 
\bq\label{4.45}
\frac{\pp L}{\pp t_k}=\{(L^{1/n})_{+}^k,L\}_{\gk}=\{L,(L^{1/n})_{-}^k\}_{\gk}
\end{equation}
where $(L^{1/n})_{+}^k=L^{1/n}*\cdots *L^{1/n}$ (k times) and here
$L=p^n+\sum_0^{n-1}u_ip^i$ with $L^{1/n}=p+\sum a_ip^{-i}$ is the $n^{th}$ root ${\bf
(D19)}\,\, L=L^{1/n}*\cdots *L^{1/n}$ (n times).  By definition of the Moyal bracket
the highest order in p on the right side of \eqref{4.42} is $n-2$ so $u_{n-1}$ is
trivial in evolution equations and can be dropped in the Lax formulation (this
changes in the Hamiltonian formulation however).  The simplest example is written out
in \cite{t1}, namely n=2 with $L=p^2+u$ and $L^{1/2}=p+\sum_1^{\infty}a_ip^{-i}$ 
and 
${\bf (D20)}\,\,
a_1=u/2;\,\,a_3=-u^2/8;\,\,a_5=u^3/16+(\gk^2/8)(u^2_x-2uu_x);\cdots$
and the first few Lax flows are
${\bf (D21)}\,\,
u_{t_1}=u_x;\,\,u_{t_3}=(3/2)uu_x+\gk^2u_{xxx};\cdots$.
This set of equations forms a Moyal KdV hierarchy which can also be obtained from
reduction of the Moyal KP hierarchy (cf. \cite{ch,c1} or noncommutative zero
curvature equations (cf. \cite{k1}).  When $\gk=0$ all higher order derivative terms
disappear and the Moyal KdV reduces to the dispersionless KdV hierarchy (cf.
\cite{ch}).  In this sense the Moyal parameter $\gk$ characterizes the dispersion
effect.  On the other hand when $\gk=1/2$ the Moyal KdV hierarchy is the ordinary KdV
hierarchy and this is due to an isomorphism of Moyal KP to ordinary KP at
$\gk=1/2$ (cf. \cite{ch,g1,s2} and Section 3).  
$\hfill\bs$
\\[3mm]\indent
We concentrate now on Examples 4.1 and 4.3 and for simplicity begin with KdV.  The
theme is somehow to exhibit the quantum mechanical nature of KdV via the dynamical
system \eqref{4.23} or \eqref{4.43}.  The momentum algebra as in this section which
leads to \eqref{4.43} and {\bf (D16)} is very attractive and we spell out more
details. We go first to \cite{t1} and extract some specific formulas to enable the
study of 
${\bf (D22)}\,\,\pp_kL=\{L_{+}^{(2n+1)/2},L\}_{\gk}$ ($t_k\to -t_k$ in {\bf (D16)}).
We will not emphasize the bihamiltonian structure.  Then for $L=p^2+u$ one has 
${\bf (D23)}\,\,L^{1/2}=p+\sum_1^{\infty}a_ip^{-i}$ with (cf. {\bf (D12)})
\bq\label{4.46}
L^{3/2}=L\star L^{1/2}=(p^2+u)\star (p+\sum_1^{\infty}a_ip^{-i})=p^3+p^2\star a_1p^{-1}
+p^2\star a_2p^{-2}+\cdots +
\end{equation}
$$+u\star p+u\star a_1p^{-1}+u\star a_2p^{-2}+\cdots$$
From \cite{t1} we have $a_{2k}=0$ and (cf. {\bf (D20)})
\bq\label{4.47}
a_1=\frac{1}{2}u;\,\,a_3=-\frac{1}{8}u^2;\,\,a_5=\frac{1}{16}u^3+
\frac{1}{8}\gk^2(u_x^2-2uu_{xx});
\end{equation}
$$a_7=-\frac{5}{128}u^4-\frac{5}{16}\gk^2(uu_x^2-2u^2u_{xx})-\frac{1}{8}\gk^2
(u_{xx}^2-2u_xu_{xxx}+2uu^{(4)});\cdots$$
The first few Lax flows are given by
\bq\label{4.48}
u_{t_1}=u_x;\,\,u_{t_3}=\frac{3}{2}uu_x+\gk^2u_{xxx};\,\,u_{t_5}=\frac{15}{8}u^2u_x
+\frac{5}{2}\gk^2(uu_{xxx}+2u_xu_{xx})+\gk^4u^{(5)};\cdots
\end{equation}
The comparison here is with {\bf (D12)} (as changed from \cite{d1}).  Thus the only 
plus terms involve the composition (since $a_2=0$)
\bq\label{4.49}
p^2\star a_1p^{-1}=a_1p+2\gk\pp_xa_1+\gk^2p^{-1}\pp_x^2a_1
\end{equation}
so ${\bf (D24)}\,\,L_{+}^{3/2}=p^3+a_1p+2\gk\pp_xa_1+u\star p=p^3+(1/2)up+\gk u_x
+u\star p$.  We know that $u\star p=pu-\gk u_x$ from calculations after \eqref{4.37}
so ${\bf (D25)}\,\,L_{+}^{3/2}=p^3+(1/2)[u\star p+\gk u_x]+\gk u_x+u\star p=
p^3+(3/2)u\star p+(3/2)\gk u_x$.
\\[3mm]\indent
{\bf REMARK 5.2.}
We see via Examples 4.1 and 4.3 that KP or KdV for example have a distinctly quantum
mechanical flavor via the phase space dynamics \eqref{4.43} (note again that it
doesn't matter whether we take $\{L_{+}^{\ga},L\}$ or $\{L,L_{+}^{\ga}\}$ to express
the time variation since it is simply a matter of time reversal).  Thinking of KdV
for simplicity, as a result one say that all $\pp u/\pp t_{2n+1}$ aarise from QM
equations
${\bf (D26)}\,\,\pp_{2n+1}=\{L,L_{+}^{(2n+1)/2}\}$.  Further, given the dependence of
the $a_i$ in {\bf (D20)} or \eqref{4.47} on $u$ one has a direct computation for 
$\pp a_{2i+1}/\pp t_{2n+1}$ (this also results from {\bf (D26)} upon writing out
terms).  The QM nature of the flows arises explicitly via the presence of $\gk$ in the
equations and this reduces to ``classical" behavior when $\gk\to 0$ (where classical
here is unrealistic since it ignores dispersion however).  Thus dispersion for example
appears as a quantum phenomenon; QM represents a smoothing or calming factor which
eliminates some wave caustics, breaking, etc.  Since one knows also that $q\sim
exp(\gk)$ is characteristic of the emergence of $q$ in various q-versions of quantum
phenomena (cf. \cite{c1}) we have here another physical motivation for $q$ which
makes the study of q-QM seem more meaningful (one can ask here also about whether
a $\gk$ quantization as in Moyal is just a first order version of a $q=exp(\gk)$
$\ul{discrete}$ quantum theory). 
On the other hand water waves for
example are eminently classical macro-phenomena so one can ask why they should obey
QM rules of behavior.  As an aside we mention, following \cite{d1}, that upon
constructing the bihamiltonian theory associated with {\bf (D26)} one finds that
$\gk$ is directly related to conformal field theory (CFT) and to the central charge
of the second Hamiltonian structure of KdV (which in turn is related to a standard
Virasoro  algebra - not q-Virasoro - cf. \cite{ch,c12,f6}).  If one thinks of KdV
(or KP) then as a QM extension of dKdV (or dKP) then some real world macro-phenomena
are essentially quantum mechanical in nature.  This is perhaps a viewpoint to be
further explored both philosophically and technically.  Note that (elementary)
particles have only one QM time evolution to drive them but in some sense ``fluids"
seem to have many, whose nature could be further examined with the study of higher
order evolutions in KdV for example.$\hfill\bs$
\begin{example}
Following \cite{d1,t1} we compute the first terms of the ``quantum" equation
$\pp_tL=\{L^{3/2}_{+},L\}_{\gk}$ with $L^{3/2}=p^3+(3/2)u\star p+(3/2)\gk u'$
(cf. {\bf (D12)} and {\bf (D25)}).  Thus first we get (since $u\star u'=uu'$)
\bq\label{4.50}
\gG=(p^3+\frac{3}{2}u\star p+\frac{3}{2}\gk u')\star (p^2+u)-(p^2+u)\star
(p^3+\frac{3}{2}u\star p+\frac{3}{2}\gk u')=
\end{equation}
$$=p^3\star u+\frac{3}{2}u\star p^3+\frac{3}{2}u\star p\star u+\frac{3}{2}\gk u'\star
p^2-[\frac{3}{2}p^2\star (u\star p)+\frac{3}{2}p^2\star\gk u'+u\star
p^3+\frac{3}{2}u\star u\star p]$$
Note now from \eqref{4.37} that ${\bf (D27)}\,\,p^3\star u=\gk^3u'''+3\gk^2pu''
+3\gk p^2u'+p^3u$ while ${\bf (D28)}\,\,u\star p^3=-\gk^3u'''+3\gk^2p-3\gk
p^2u'+p^3$. Hence ${\bf (D29)}\,\,p^3\star u-u\star p^3=2\gk^3u'''+6\gk p^2u'$. 
Next consider
$\Xi=(3/2)(u\star p^3-p^2\star u\star p)$ with $u\star p=p\star u-2\gk u'$ so
${\bf (D30)}\,\,\Xi=(3/2)(u\star p^3-p^3\star u+2\gk p^2\star
u')=-(3/2)(2\gk^3u'''+6\gk p^2u')+(3/2)2\gk p^2\star u'=-9\gk p^2u'+3\gk(p^2u'
+2\gk pu'')=-6\gk p^2u'+6\gk pu''$.  Finally ${\bf (D31)}\,\,(3/2)(u\star p\star
u-u\star u\star p)=(3/2)(u\star (u\star p+2\gk u')-u\star u\star p)=3\gk u\star u$ and
${\bf (D32)}\,\,(3/2)\gk(u'\star p^2-p^2\star u')=-6\gk pu''$.  Combining then
{\bf D27)} - {\bf (D32)} we get
${\bf (D33)}\,\,\gG=2\gk^3u'''+3\gk u\star u'$ so the Lax equation then decrees that
${\bf (D34)}\,\,u_t=(1/2\gk)(2\gk^2 u'''+3\gk uu')=
(3/2)uu'+\gk u'''$ which is the standard KdV form with the dispersion coming from $\gk$. 
$\hfill\bs$
\end{example}

\section{REMARKS ON Q-THEORIES}
\renewcommand{\theequation}{6.\arabic{equation}}
\setcounter{equation}{0}

QKP (and qKdV) can be developed in a hierarchy form following \cite{a2,a3,a4,c10,
f2,h3,i1,k8,t3,t4}. 
This is pursued in connection with Hirota formulas at some length in \cite{c10}
and we remark here only that the resulting qKP or qKdV equations are very
complicated due to formulas of the form
\bq\label{23}
u=a_1=(1-D)\left(\frac{(1/2)(\pp_1^2-\pp_2)\tau_q}{\tau_q}\right)
-\left(\frac{\pp_1
\tau_q}{\tau_q}\right)^2+\frac{\pp_1\tau_q}{\tau_q}D\left(\frac{\pp_1\tau_q}
{\tau_q}\right)+D_q\frac{\pp_1\tau_q}{\tau_q}
\end{equation}
instead of the classical $u=\pp^2log(\tau)$.  Similarly for qKdV one has a difficult
formula ${\bf (A31)}\,\,u=\pp_q\pp_1log\tau(x,t)D\tau(x,t)$.
\\[3mm]\indent
{\bf REMARK 6.1.}
In this direction if one actually writes out a qKdV equation for example from the
hierarchy picture it will have the form (cf. \cite{c10})
\bq\label{24}
\pp_tu=
(\pp_q^3u)+w_2(\pp_q^2u)+w_1(\pp_qu)-[(\pp_q^2w_0)+u_1(\pp_qw_0)]
\end{equation}
where (setting ${\bf (E1)}\,\,u_1=(q-1)xu=(1+D)s_0$ and $u=s_1+Ds_1+s_0^2+\pp_qs_0$)
\bq\label{25}
w_2=D^2s_0+u_1=D^2s_0+Ds_0+s_0;
\end{equation}
$$w_1=(q+1)(D\pp_qs_0)+D^2s_1+[(Ds_0)+s_0](Ds_0)+
u;$$
$$w_0=\pp_q^2s_0
+(q+1)(D\pp_qs_1)+u_1\pp_qs_0+u_1(Ds_1)+us_0+D^2s_2$$
(cf. \cite{c1,c11,d17} for other forms
of q-equations or noncommutative integrable equations).
It appears therefore that after expressing e.g. $s_0=(1+D)^{-1}(q-1)xu$ with ${\bf 
(E2)}\,\,(1+D)^{-1}\sim\sum_0^{\infty}(-D)^n$ formally the qKdV equation will have an
infinite number of terms.  A similar kind of equation for ``qKdV" was derived in
\cite{c12} (cf. also \cite{c16}), by use of a version of q-Virasoro, in the form
${\bf (E3)}\,\,u_t+c\pp_q^2(D+D^{-1})^{-1}\pp_qu+\pp_q(uDu)+D^{-1}u\pp_qD^{-1}u=0$
where here $\pp_qf=[f(qx)-f(q^{-1}x)]/(q-q^{-1})x$.$\hfill\bs$
\\[3mm]\indent
One recalls here that 
$\pp_qf(z)=[f(qz)-f(z)]/ (q-1)z$ with ${\bf
(E4)}\,\,\pp_q(fg)=\pp_q(f)g+\tau(f)\pp_qg$ where
$\tau(f)(z)=f(qz)$ (we use $\tau$ and D interchangably now and note that
$\pp_q\tau=q\tau\pp_q$).  PSDO are defined via  an equation
${\bf (E5)}\,\,A(x,\pp_q)=\sum_{-\infty}^nu_i(x)\pp^i_q$ with $\pp_qu=
(\pp_q u)+\tau(u)\pp_q$ and one has ($D_q\sim\pp_q$)
\bq\label{3.22}
\pp_q^{-1}u=\sum_{k\geq 0}(-1)^kq^{-k(k+1)/2}(\tau^{-k-1}(\pp_q^ku))\pp_q^{-k-1};\,\,
\pp_q^nu=\sum_{k\geq 0}\left[\begin{array}{c}
n\\
k\end{array}\right]_q(\tau^{n-k}(\pp_q^ku))\pp_q^{n-k}
\end{equation}
Recall also 
\bq\label{3.23}
(n)_q=\frac{q^n-1}{q-1};\,\,\left[\begin{array}{c}
m\\
k\end{array}\right]_q=\frac{(m)_q(m-1)_q\cdots(m-k+1)_q}{(1)_q(2)_q\cdots (k)_q}
\end{equation}
There are then q-analogues of the Leibnitz rule etc. and for
$L_q=\pp_q+u_1(z)+u_2(z)\pp_q^{-1}+ u_3(z)\pp_q^{-2}+\cdots$ and one has q-KP via
$(\pp L_q/\pp t_m)=[(L_q^m)_{+},L_q]$ where
$u_1(z)$ has a nontrivial evolution. 
In accord with Section 3 we should now represent the ring or
algebra
$\mf{A}_q$ of qPSDO symbols via a product as in say {\bf (B7)} and thence provide 
expressions for deformation thereof.  The $X$ and $P$ variables should come from the
phase space for dKP.  Evidently the qPSDO symbols will involve a variation on
\eqref{3.21} and one can utilize techniques of \cite{a2,a3,a4,c1,m4,k8} (cf. also
\cite{c2,c10,c11} for q-formulas) for calculations.  In this direction
one finds (heuristically) that the
ring or algebra calculi of ${\mf A}\,\sim$ PSDO and ${\mf A}_q$ correspond
symbolically via
$\pp\sim D_q=\pp_q$ and suitable insertion of $D\sim \tau$ factors along with
q-subscripts; in particular ${\bf (E6)}\,\,(\pp_{\xi})_q^k\xi^i=i_q\cdots (i-k+1)_q
(\tau\xi)^{i-k}$ is needed.  
When commutators are also envisioned individual
terms may differ because e.g. brackets
$[\,\,,\,\,]$ have different degrees, etc. but one notes that e.g.
\bq\label{3.30}
[a\pp,b\pp]=(ab'-ba')\pp;\,\,[a\pp_q,b\pp_q]=(ab'_q-ba'_q)\pp_q+(a\tau b-b\tau a)\pp_q^2;
\end{equation}
$$(ab'-ba')\pp\to (a\pp_qb-b\pp_qa)\pp_q=[a(b'_q+\tau b\pp_q)-b(a'_q+\tau
a\pp_q)]\pp_q$$
\\[3mm]\indent\indent
{\bf REMARK 6.2.}
Given $L_q=D+q+b_0(t)+\sum_1^{\infty}b_{-i}D_q^{-i}$ for qKP and
$L=\pp+\sum_1^{\infty}\gb_{-i}\pp^{-i}$ for KP, with nontrivial evolution of $b_0$ one
cannot perhaps expect KP and qKP to be ``isomorphic" but 
as indicated above they do correspond symbolically so there seems to be no reason to
regard one as more ``quantum" than the other (except perhaps the fact that the
q-theory represents a discretized version of the other and such discretization may be
an essential quantum signature - not yet authenticated in conventional QM such as the
Moyal quantization).   There is also another way to view qKP due to E. Frenkel (cf.
\cite{f2}) via 
${\bf (E7)}\,\,\widetilde{qKP}:\,\,\tl{L}_q=D+a_0(t)+\sum_1^{\infty}a_{-i}(t)D^{-i}$
where
$Df(t)=f(qt)$ (recall here that $(q-1)xD_qf=(D-1)f$).  It is shown in e.g.
\cite{a2,a3} (cf. also
\cite{c1}) that there is an isomorphism mapping qKP or $\widetilde{qKP}$ into the
discrete KP hierarchy which exhibits their equivalence via
\bq\label{3.39}
a_i(y)=\sum_{0\leq k\leq n-i}\frac{\left[\begin{array}{c}
k+i\\
k\end{array}\right]}{(-y(q-1)q^i)^k}b_{k+i}(y)
\end{equation}
We refer to \cite{a2,a3} for details on discrete KP which in fact is equivalent to
the 1-Toda lattice (cf. \cite{a2,a2}).  The correspondence to qKP can be seen best
through the notation of \cite{t2} where one works with Toda Lax operators $L=\gL
+\sum_0^{\infty}u_{n+1}(\gep,t,\bar{t},s)\gL^{-n}$ with $\gL\sim exp(\gep\pp_s)$ 
(cf. also \cite{ch}).
We note also connections of discrete KP to algebraic equations over finite fields in
\cite{b7} and an interesting paper \cite{d3} on formulating quantum mechanics with
difference operators.
$\hfill\bs$

%\newpage

\end{document}